\documentclass[]{interact}

\usepackage[caption=false]{subfig}

\usepackage{xcolor}
\usepackage{lmodern}

\usepackage[numbers,sort&compress,merge]{natbib}
\bibpunct[, ]{[}{]}{,}{n}{,}{,}

\theoremstyle{plain}

\theoremstyle{definition}

\theoremstyle{remark}

\begin{document}

\articletype{Research article}

\title{Evaporation Characteristics of Heat Pipes with Sub-Critical Nanopores}

\author{
\name{Sumith Yesudasan\thanks{CONTACT: Sumith Yesudasan. Email: syesudasan@newhaven.edu}}
\affil{Mechanical and Industrial Engineering, University of New Haven, CT, USA}
}

\maketitle

\begin{abstract}
This study explores heat transfer mechanisms in heat pipes with sub-critical nanopores using coarse-grained molecular dynamics (CGMD) simulations, aiming to enhance thermal management in nanoscale applications. With the increasing need for efficient cooling solutions in microelectronics and high-performance computing, nanoporous heat pipes have gained attention due to their high thermal conductivity and passive operation. This research evaluates the effects of pore size, temperature gradients, and water fill ratios on the heat transfer efficiency of heat pipes with 2 nm and 3 nm diameter nanopores. The results indicate that larger temperature gradients significantly enhance heat transfer rates, while filled heat pipes perform better than medium-filled ones, primarily due to more effective phase change and fluid flow dynamics. Notably, the 2 nm filled models show improved performance over the 3 nm models, suggesting an optimal balance between capillary action and fluid resistance at smaller pore sizes. The study also reveals that in sub-critical nanopores, surface-driven flows are more effective than traditional wicking, underscoring the role of surface interactions in optimizing heat transfer. These findings provide critical insights for designing and optimizing nanoporous heat pipes, offering practical guidance for developing more efficient thermal management systems in electronics cooling and other applications.
\end{abstract}


\section{Introduction}

The demand for efficient thermal management solutions has surged with advancements in microelectronics and high-performance computing systems \cite{garimella2016electronics, hu2020emerging, khalaj2017review, reed2022reinventing, shalf2020future}. Heat pipes, which are known for their high thermal conductivity and passive operation, play a crucial role in thermal management applications. Traditional heat pipes rely on capillary action within wick structures, but recent innovations in nanotechnology have led to the development of nanoporous structures with enhanced heat transfer capabilities \cite{guo2020enhancement, tang2023review, yi2023experimental}. These advancements facilitate the application of heat pipes in micro and nanoscale devices, including thinner cell phones, sleeker batteries, and lighter laptops.

Studying evaporation mechanism and heat transfer in heat pipes traditionally involves coupled Navier-Stokes and heat transfer equations \cite{alihosseini2021experimental, pawar2020cfd, behi2020thermal}. However, these continuum-level techniques are insufficient at many nanoscale systems \cite{daivis2018challenges, liu2011validity, alami2023critical} where the bulk property of the system cannot be represented using continuous functions. Molecular dynamics simulations, particularly classical Newtonian-based approaches, offer an alternative but are computationally expensive due to long-range Coulombic forces and detailed hydrogen atom modeling in water molecules. This challenge can be addressed by using coarse-grained molecular dynamics (CGMD), a technique that simplifies the model by aggregating multiple water molecules into single spherical beads interacting through a force field that statistically reproduces water's thermodynamic properties \cite{mccabe2012improved, ouyang2015thermodynamic, molinero2009water}.

Previous research on water evaporation using classical molecular dynamics (MD) simulations was often limited by high computational costs. To overcome these challenges, in a prior study, the author of this paper developed coarse-grained molecular dynamics (CGMD) models of water utilizing the Morse potential. These models were specifically designed to investigate evaporation from hydrophilic nanopores with diameters ranging from 2 to 5 nm \cite{yesudasan2022critical}. The study found that the \textbf{critical diameter} for sustained, continuous evaporation in nanopores is between 3 and 4 nm. Building on these insights, the current study delves into the heat and mass transfer dynamics within conceptual nanoscale heat pipes featuring sub-critical (below critical diameter) nanopores, using CGMD simulations. The objective is to understand the impact of capillary wicking action in these heat pipes when sub-critical nanopores are present.

In conventional heat pipes, heat is transported from the evaporator section to the condenser through water vapor, with the condensed water being returned via a porous wick structure. This wick structure, typically porous, facilitates water transport through capillary action, making it particularly suitable for low-gravity applications. The capillary pressure (\(P_c\)) generated within a wick structure is inversely proportional to the pore radius (\(r\)) of the wick material, as described by the Young-Laplace equation:

\begin{equation}
    P_c = \frac{2 \gamma \cos(\theta)}{r}
    \label{equan1}
\end{equation}

According to this equation, as the radius of the wicking pores decreases, the capillary pressure increases, thereby enhancing the wicking phenomenon. However, reducing the pore radius also increases surface tension ($\gamma$) and decreases the permeability of the wick \cite{israelachvili2011intermolecular}. Therefore, an optimal range of pore sizes is needed to maximize wicking efficiency. The critical question is: \textbf{How small should these pores be to maximize the benefits of wicking?} The current study aims to explore systems with pore diameters of 2 nm and 3 nm, providing insights into the behavior of heat pipes incorporating sub-critical nanopores.

Two heat pipe models with different nanopore diameters (2 nm and 3 nm) are analyzed under varying thermal and water fill conditions. The study investigates the effects of temperature gradients and fill ratios on the thermal performance of heat pipes. By comparing models with different water content levels—specifically "filled" and "medium fill" conditions—the research provides insights into the impact of fluid dynamics on heat transfer efficiency.

The study reveals that heat transfer in heat pipes with sub-critical nanopores is predominantly influenced by surface interactions (flow along the internal surface of the heat pipe, instead of flowing through the porous structure). The results emphasize that, when aiming to enhance capillary wicking, it is crucial to consider intermolecular forces, which can significantly influence the transport dynamics. To the best of our knowledge, this study represents the first molecular-level investigation of transport properties in wicks with sub-critical nanopores under a thermal gradient.

\section{System and simulation setup}

\begin{figure}[!h]
\centering
\includegraphics[width=0.65\textwidth]{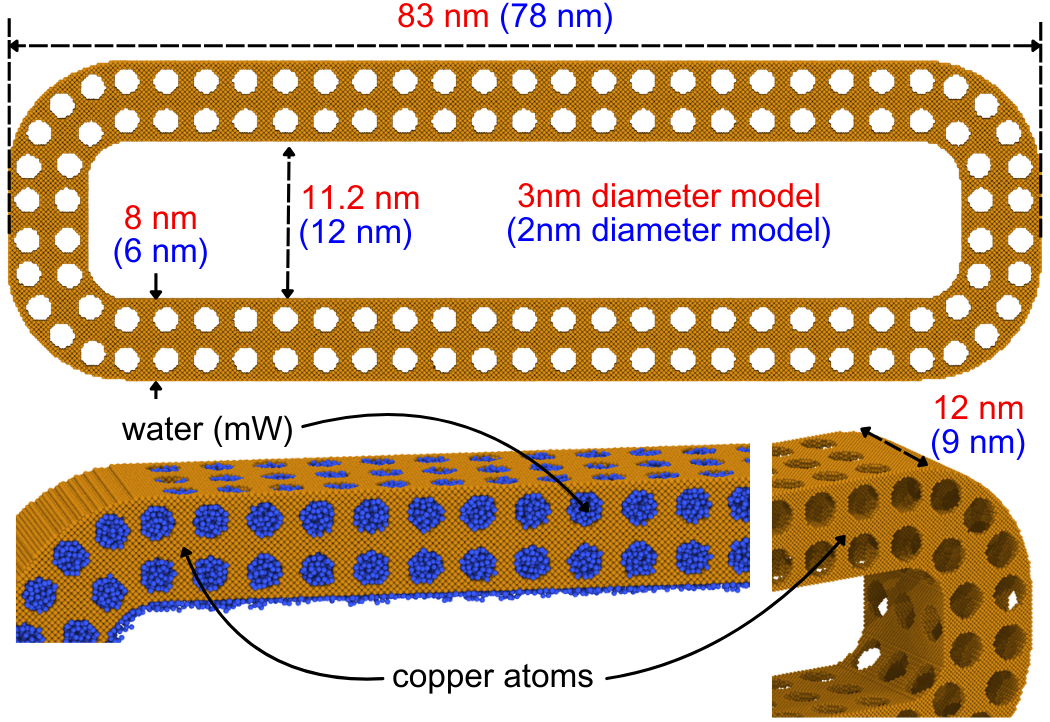}
\caption{The top panel shows boundary conditions and dimensions of the molecular systems used in this study, using the 2 nm copper model as a guide. The dimensions of the 2 nm model are labeled in {\color{blue} blue}, and those of the 3 nm model are shown in {\color{red} red}. The bottom panels show the images of the system with water and without water.}
\label{figure1}
\end{figure}

To simulate the heat pipe at the nanoscale, two systems with varying levels of water inside are considered. The first system is a 3D porous heat pipe with 2 nm diameter holes drilled through it in all directions except at the corners, as shown in Fig. \ref{figure1}. The length of the heat pipe is 78 nm, its thickness is 6 nm, and the gap inside is 12 nm. This gap is necessary for the vapor to travel to the cold side and condense. 

The second system considered for the study is a pipe with 3 nm diameter holes, a length of 83 nm, a thickness of 8 nm, and a gap of 11.2 nm. The dimensions of both systems are labeled and shown in Fig. \ref{figure1}. For clarity, only the image of the 2 nm system (henceforth called the 2 nm model) is shown, with labels in blue for the 2 nm model and red for the 3 nm model. The widths of the models are 9 nm and 12 nm, respectively. For brevity, only first system is shown and dimensions of second system is overlaid above it.

\begin{figure}[!h]
\centering
\includegraphics[width=0.65\textwidth]{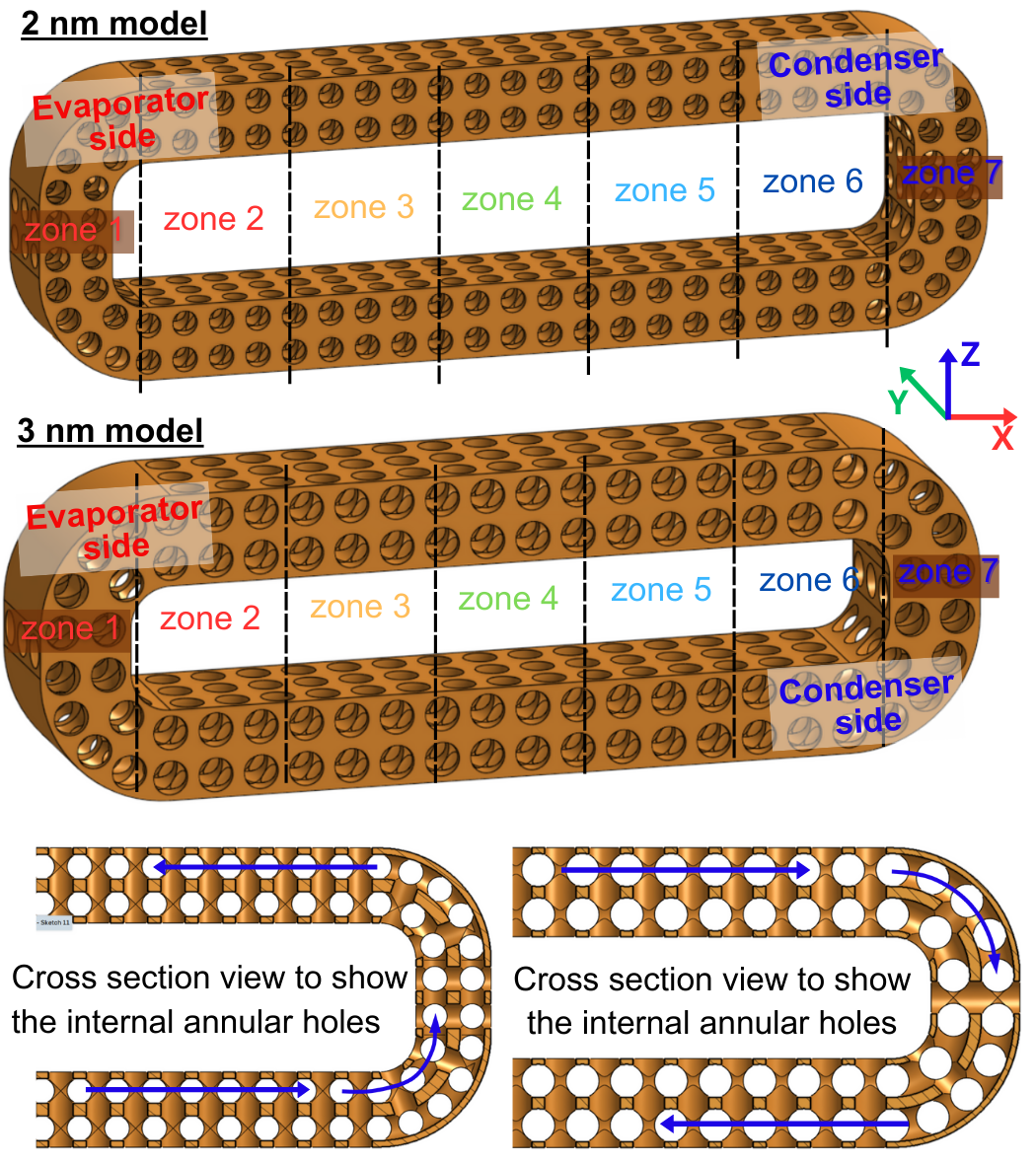}
\caption{The 3D model images of both the 2 nm and 3 nm models are displayed in the top and center, respectively. Holes are drilled in the x, y, and z directions, except at the curved corners. At these curved corners, connectivity is maintained by linking the holes drilled along the x-axis and z-axis. The cross-sectional views shown in the bottom right and left provide a better understanding of this connectivity. The arrows in these views illustrate the connectivity and are not intended to represent the fluid flow path.}
\label{fig_temp_zones}
\end{figure}

The boundaries of the 2 nm model and the 3 nm model are reflective on the top, bottom, and sides, which contain the water molecules within the system boundaries. However, along the axis normal to the image, the boundary is periodic. The corners of the copper tube are covered with copper atoms with no holes to prevent water leakage into the empty space at the corners. The water molecules are shown in blue, and the copper atoms are shown in reddish-brown. The molecular details, especially the force field, will be discussed in the next section.

Figure \ref{fig_temp_zones} illustrates the three-dimensional (3D) models of the 2 nm and 3 nm systems in an isometric perspective. These models represent the spatial arrangement of copper atoms within the system. To clarify the annular connectivity between the holes in both models, cross-sectional views are provided in the bottom panels of Fig. \ref{fig_temp_zones}. The axes of the geometry are depicted for reference. The arrows in the cross-sectional views serve to demonstrate the connectivity within the system, rather than indicating the direction of water flow. The actual direction of water flow will be analyzed and detailed in the Results and Discussion section.

\subsection{Temperature zoning}
To control the thermal conditions of the molecular system, the simulation domain is divided into distinct temperature zones along the x-axis. As illustrated in Fig. \ref{fig_temp_zones}, there are seven zones, with Zone 1 representing the hottest region and Zone 7 representing the coldest. Both Zone 1 and Zone 7 extend 10 nm from their respective ends of the simulation box, while the remaining zones (Zone 2 through Zone 6) are equidistantly spaced between them. The temperature gradient is established across these zones to achieve a controlled thermal environment within the system.

In this setup, water molecules are not confined to a single zone; instead, they move freely and frequently cross between zones due to their inherent kinetic energy and thermal motion. To accurately represent the system's thermal behavior, the spatial distribution of water molecules is continuously monitored and updated at each simulation step. This process ensures that any changes in the molecular population within each zone are accounted for, preventing discrepancies in the temperature distribution.

Temperature values for Zones 2 through 6 are determined through linear interpolation between the temperatures of Zone 1 and Zone 7. For instance, when the target temperature for Zone 1 is set to 400 K, a temperature gradient is established, gradually decreasing to 300 K in Zone 7. In another scenario, if Zone 1 is maintained at 350 K, the temperature again tapers down to 300 K in Zone 7. This gradient creates a smooth thermal transition across the simulation domain.

In our simulations, temperatures are applied exclusively to the water molecules through a defined thermal gradient along the x-axis, ranging from the hot end to the cold end of the heat pipe. The copper atoms constituting the heat pipe walls are not directly thermostatted; instead, they interact thermally with the water molecules via molecular interactions. This approach focuses on capturing the molecular-level heat transfer within the working fluid, particularly within the sub-critical nanopores. By applying the temperature gradient to the water molecules, we aim to study the effects of thermal energy input and extraction on the fluid dynamics and heat transfer mechanisms, while acknowledging that the copper walls serve primarily as structural boundaries in this model. The simulations employed the Nosé-Hoover NVT thermostat. Specifically, a modified extension developed by Martyna, Tobias, and Klein \cite{martyna1992nose} was used, which incorporates an additional thermostat chain to enhance ergodicity.

The dynamic nature of the molecular system means that water molecules are constantly migrating between these defined zones. As a result, the temperatures and molecular compositions of each zone must be updated regularly during the simulation to reflect the actual state of the system. Although this process is computationally demanding, it is crucial for maintaining the accuracy and reliability of the simulation results. Continuous updating ensures that the thermal environment is accurately modeled, which is vital for studying temperature-dependent molecular behaviors and interactions.

\subsection{Molecular modeling details}

For the simulation of water molecules, instead of using the traditional computationally expensive models, a popular coarse grained model  called mW model\cite{molinero2009water} is used. This model represents one water molecule by a coarse grained bead. The traditional models of water utilize long-range forces to replicate the hydrogen-bonded structure of water. The mW model diverges from this approach by introducing a short-range, angular-dependent term that promotes tetrahedrality. This model successfully reproduces the density, structure, and various phase transitions of water with high accuracy, at a significantly reduced computational cost compared to atomistic models. The mW model accurately reproduces water's density maximum, melting temperature, and enthalpy of vaporization, among other properties. The results show that the structural and thermodynamic behavior of water can be effectively modeled by focusing on the tetrahedral connectivity of the molecules, rather than the nature of the interactions.

The equations governing this mW potential is based on the Stillinger-Weber model and is given in the below equations.

\begin{equation}
    E_{mW-mW} = \sum_{i} \sum_{j>i} \phi_2(r_{ij}) + \sum_{i} \sum_{j \neq i} \sum_{k > j} \phi_3(r_{ij}, r_{ik}, \theta_{ijk})
\end{equation}

\begin{equation}
    \phi_2(r_{ij}) = A \epsilon \left[ B \left( \frac{\sigma}{r_{ij}} \right)^m - \left( \frac{\sigma}{r_{ij}} \right)^n \right] \exp \left( \frac{\sigma}{r_{ij} - a\sigma} \right)
    \label{eq_2body}
\end{equation}

\begin{align}
    \nonumber \phi_3(r_{ij}, r_{ik}, \theta_{ijk}) = \lambda \epsilon \left( \cos \theta_{ijk} - \cos \theta_0 \right)^2  \\
    \exp \left( \frac{\sigma}{r_{ij} - a\sigma} \right) \exp \left( \frac{\sigma}{r_{ik} - a\sigma} \right)
    \label{eq_3body}
\end{align}

The mW potential is sum of a two body interaction potential Eq. \eqref{eq_2body} and a three body interaction potential Eq. \eqref{eq_3body}. The subscripts $i, j, k$ represents the three bodies (atoms/molecules) under consideration. The detailed explanation of this potential and relevance of each term is beyond the scope of this paper and readers are advised to refer the original work by Molinero \cite{molinero2009water}.

The copper atoms are arranged in a Face-Centered Cubic (FCC) lattice structure with a lattice constant of 3.615 Å, reflecting the atomic configuration of copper. Porous nanostructures are generated by systematically removing material to create holes with diameters of 2 nm and 3 nm, oriented vertically and horizontally, using custom-developed computer code. These holes are interconnected by drilling annular channels, mimicking the structural design of a heat pipe. In molecular dynamics (MD) simulations, copper atoms are typically modeled using the Embedded Atom Model (EAM) potential, which has been well-established for simulating metallic interactions due to its effective representation of atomic bonding and coordination effects \cite{zarringhalam2019effects, lysogorskiy2021performant, fairushin2020numerical, cometto2021copper}. However, the EAM potential for copper becomes unstable at larger time steps, such as 10 fs, whereas it is optimized for smaller time steps around 1 fs. To address this instability, copper-copper interactions were excluded from the time integration scheme, allowing copper atoms to interact exclusively with water molecules through the Lennard-Jones potential. Additionally, to keep the copper atoms in place, a harmonic spring algorithm is used with a spring constant of 10. This selective interaction modeling ensures that the static copper framework minimally affects the dynamic behavior of the coarse-grained water models, thus maintaining the accuracy of the fluid dynamics in proximity to the copper.

The interaction between water and copper is modeled using the Lennard-Jones potential, described by Eq. \eqref{eqn_LJ}:

\begin{equation}
    E_{mW-Cu} = 4\epsilon \left[ \left( \frac{\sigma}{r} \right)^{12} - \left( \frac{\sigma}{r} \right)^6 \right]
    \label{eqn_LJ}
\end{equation}

The parameters for this potential are calibrated to simulate a hydrophilic interaction between water and copper, consistent with findings reported by Huang et al. \cite{huang2023critical}. For this equation, values for parameters are $\epsilon$ = 0.0342 kcal/mol and $\sigma$ = 0.28675 nm.

\begin{figure*}[!h]
\centering
\includegraphics[width=0.6\textwidth]{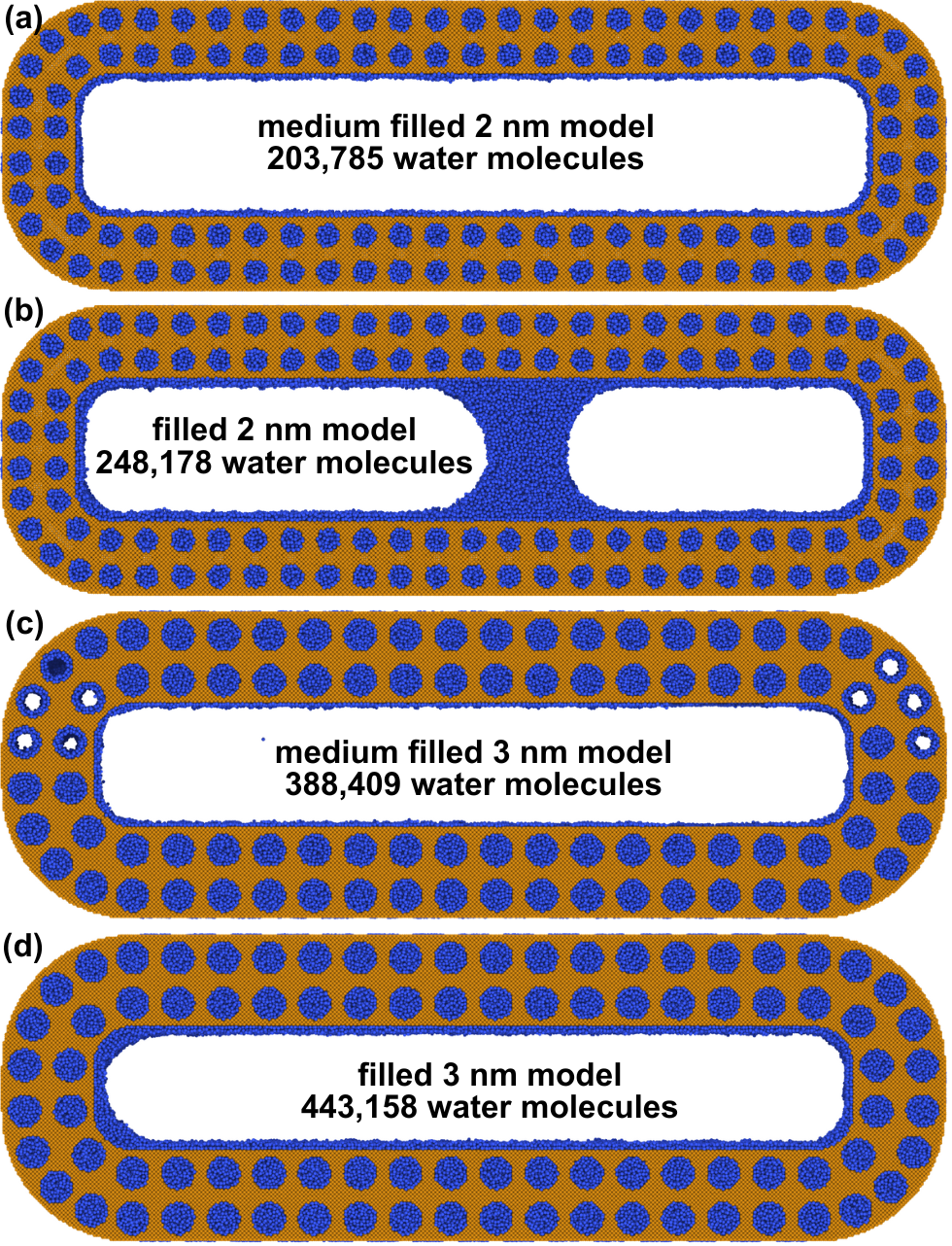}
\caption{Equilibrated molecular models of the heat pipe with 2 nm holes filled with (a) minimal water and (b) excess water, and 3 nm holes filled with (c) minimal water and (d) excess water.}
\label{figure3}
\end{figure*}

To examine the influence of water quantity on the system and its effect on circulation dynamics, two levels of water content were considered as shown in the Fig. \ref{figure3}. The "filled" systems, characterized by an excess amount of water, contain 248,178 water molecules for the 2 nm hole model and 443,158 water molecules for the 3 nm hole model. Conversely, the "medfill" models, which represent the minimum water required to adequately wet the copper pipe, contain 203,785 water molecules for the 2 nm model and 388,409 water molecules for the 3 nm model. The 2 nm pore model comprises 182,180 copper atoms, while the 3 nm pore model consists of 391,660 copper atoms. This dual approach allows for a comprehensive understanding of how varying water volumes influence the thermal and fluid dynamics within the nanoscale heat pipe models.

\subsection{Simulation Parameters}
The simulations utilize an integration time step of 10 femtoseconds (fs) for both water and copper, ensuring numerical stability and accurate representation of atomic motions. The cutoff radius for the potential functions, specifically for copper-water interactions, is set to 1.2 nanometers (nm), reflecting the short-range nature of these interactions. The mass of copper atoms is considered as 63.546 g/mol, consistent with its atomic mass, while the coarse-grained molecular dynamics (CGMD) water beads are assigned a mass of 18.01468 g/mol, representing the effective mass of water molecules in the coarse-grained model.

Reflective boundary conditions are applied in the x and z directions, while periodic boundary conditions are imposed along the y-axis. This setup emulates a confined system with infinite extension in one direction, suitable for studying nanoscale transport phenomena. The geometric orientation and boundary conditions of the model are illustrated in Fig. \ref{fig_temp_zones}. 

For the interaction between copper and water, a neighbor list is explicitly generated, allowing for efficient computation of pairwise forces within the cutoff distance. The Stillinger-Weber (mW) potential used for water-water interactions has an internal implementation in the LAMMPS software, negating the need for an explicit cutoff definition. This built-in approach optimizes computational performance by efficiently handling short-range interactions inherent to water's molecular structure.

The simulation protocol begins with an equilibration phase lasting two million time steps to ensure system stability and thermal equilibration. This is followed by a production phase of half a million time steps, during which data is collected at regular intervals. Key properties such as velocity, temperature, and pressure are monitored and recorded every 5 steps to capture detailed dynamic behavior.

The collected simulation data is then post-processed using custom C++ code developed in-house \cite{standardcpp} and MATLAB software \cite{MATLAB}.


\section{Results and discussion}

The molecular models described in the previous sections are utilized to simulate the heat pipe. For convenience, the cases are named according to the hole diameter, followed by the amount of water and the temperature at the left end of the pipe. For example, ``2nm-medfill-400K-Density'' refers to the density plot for a model with medium water fill, 2 nm diameter nanopores, and a left end temperature of 400 K.

Initially, the heat pipe is modeled as a copper pipe, and water is introduced into it as a rectangular block. This water is absorbed by the nanopores, after which a new batch of water is introduced. The system is equilibrated for 500,000 time steps (5 ns), and this process is repeated until the pipe is completely wet. For the ``filled'' cases, the process continues until excess water accumulates in the inner region of the heat pipe.

\begin{figure*}[!h]
    \centering
    \includegraphics[width=0.5\textwidth]{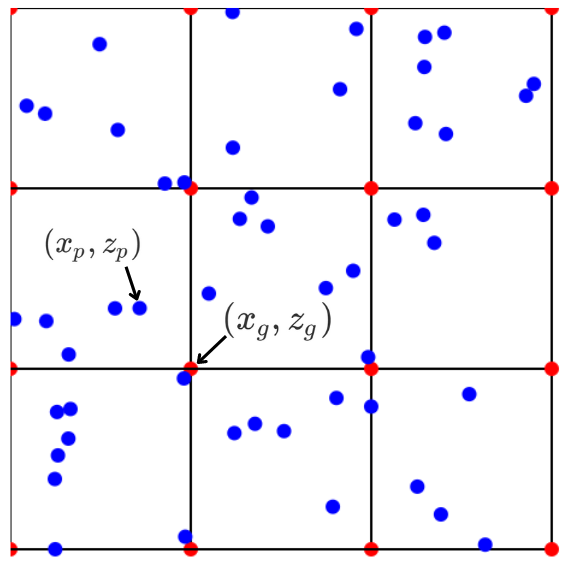}
    \caption{A sample representation of a segment of the hypothetical grid utilized for mapping molecular data to continuum-level properties is illustrated in the figure. In this representation, $(x_p, z_p)$ denotes the positions of particles (atoms), while $(x_g, z_g)$ indicates the coordinates of the grid points. This grid enables the conversion of discrete molecular information into a continuous field, helpful for the analysis of macroscopic properties.}
    \label{fig_grids}
\end{figure*}

Once equilibrated, the systems undergo production runs for at least 5 ns. During this period, a 2D grid-based approach is employed to map the statistical quantities to continuum-level properties. Figure \ref{fig_grids} illustrates a sample representation of such a grid with grid points \((x_g, z_g)\) and particles at positions \((x_p, z_p)\). The properties of interest, computed at the particle positions, are interpolated onto the grid points using Nearest Grid Point (NGP) interpolation.

\subsection{Nearest Grid Point (NGP) Interpolation}

In NGP interpolation, each particle's property is assigned to the nearest grid point, providing a computationally efficient way to map atomic data onto a continuum grid. However, this approach can introduce aliasing errors, particularly when the grid resolution is too coarse to accurately capture rapid variations in properties such as density, temperature, or velocity. The grid value at a point \((x_g, z_g)\) is updated by accumulating the properties of particles closest to that grid point. When the grid spacing is inadequate, high-frequency details are inaccurately represented as lower frequencies, leading to errors in the simulation results that can propagate through subsequent calculations, affecting the accuracy of derived properties and material behavior predictions. To address this issue, our study employs a finer grid spacing of 0.5 nm for data interpolation, ensuring that rapid variations in properties are better captured and reducing the impact of aliasing errors on the simulation outcomes.
.

Let \( f(t, x_p, z_p) \) represent the property at particle position \((x_p, z_p)\) at time \( t \). The grid value \( f(t, x_g, z_g) \) at time \( t \) is given by:

\begin{equation}
f(t, x_g, z_g) = \sum_{(x_p, z_p) \in \text{NGP}(x_g, z_g)} f(t, x_p, z_p)    
\end{equation}

where \(\text{NGP}(x_g, z_g)\) denotes the set of particles nearest to the grid point \((x_g, z_g)\). A grid spacing of 0.5 nm is used for all our cases unless mentioned explicitly.

\subsection{Temporal Averaging}

To reduce fluctuations and obtain smoother properties over time, temporal averaging is performed. Suppose we are averaging over \( n \) time steps, with the property values at time steps \( t_1, t_2, \ldots, t_n \). The temporally averaged property \(\bar{f}(x_g, z_g)\) at the grid point \((x_g, z_g)\) is:

\begin{equation}
\bar{f}(x_g, z_g) = \frac{1}{n} \sum_{k=1}^{n} f(t_k, x_g, z_g)    
\end{equation}

The following equations describe the interpolation of specific properties using NGP and temporal averaging.

\subsubsection{Velocity Components}

For the velocity components \( v_x \) and \( v_z \):

\begin{equation}
v_x(t, x_g, z_g) = \sum_{(x_p, z_p) \in \text{NGP}(x_g, z_g)} v_x(t, x_p, z_p)    
\end{equation}

\begin{equation}
v_z(t, x_g, z_g) = \sum_{(x_p, z_p) \in \text{NGP}(x_g, z_g)} v_z(t, x_p, z_p)
\end{equation}

Temporal averaging:

\begin{equation}
\bar{v}_x(x_g, z_g) = \frac{1}{n} \sum_{k=1}^{n} v_x(t_k, x_g, z_g)
\end{equation}

\begin{equation}
\bar{v}_z(x_g, z_g) = \frac{1}{n} \sum_{k=1}^{n} v_z(t_k, x_g, z_g)
\end{equation}

\subsubsection{Density}

For the density \( \rho \):

\begin{equation}
\rho(t, x_g, z_g) = \sum_{(x_p, z_p) \in \text{NGP}(x_g, z_g)} \rho(t, x_p, z_p)
\end{equation}

Temporal averaging:

\begin{equation}
\bar{\rho}(x_g, z_g) = \frac{1}{n} \sum_{k=1}^{n} \rho(t_k, x_g, z_g)
\end{equation}

\subsubsection{Temperature}

For the temperature \( T \):

\begin{equation}
T(t, x_g, z_g) = \sum_{(x_p, z_p) \in \text{NGP}(x_g, z_g)} T(t, x_p, z_p)
\end{equation}

Temporal averaging:

\begin{equation}
\bar{T}(x_g, z_g) = \frac{1}{n} \sum_{k=1}^{n} T(t_k, x_g, z_g)
\end{equation}

The properties of velocity,  density, and temperature at grid points $(x_g, z_g)$ are estimated using the Nearest Grid Point (NGP) interpolation method. This involves assigning each particle's property to the nearest grid point and performing temporal averaging over multiple time steps to smooth out fluctuations. The resulting plots, which illustrate these interpolations, are discussed in the following sections, providing a detailed explanation and analysis.

\begin{figure*}[!h]
    \centering
    \includegraphics[width=\textwidth]{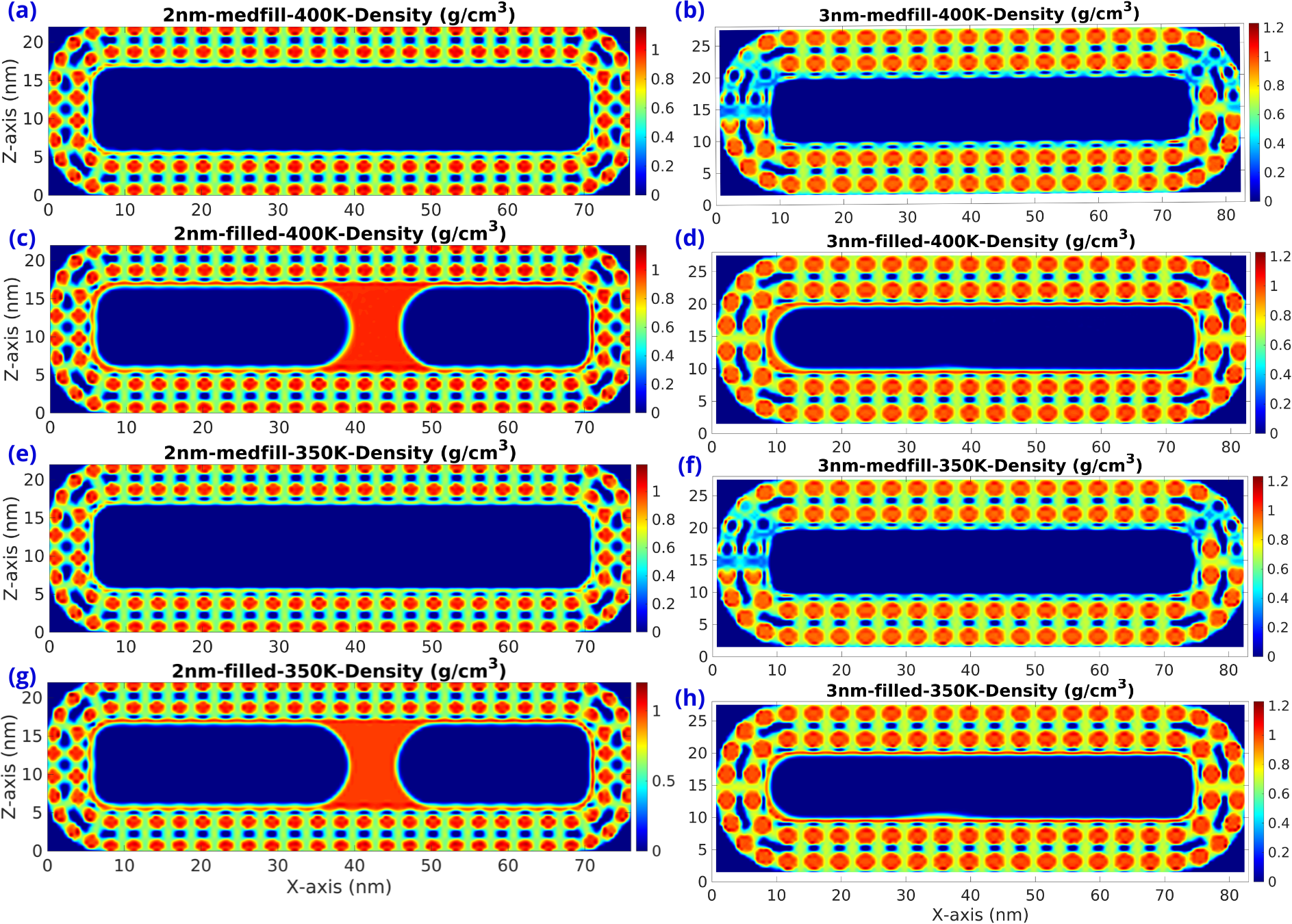}
    \caption{Density plots for 2 nm model (1st column panels) and 3 nm model (2nd column panels) are given for high temperature (400 K) case in top four panels and low temperature (350 K) case in bottom four panels. Medium filled water cases are given in panels a, b, e, and f and filled water cases are given in c, d, g and h respectively. The color scale (bar) shows the density of water in $g/cm^3$}
    \label{fig_density}
\end{figure*}

The mapped density plots for all cases are shown in Fig. \ref{fig_density}. The regions in red indicate high-density (liquid) water, while those in blue represent water vapor or vacuum. Areas with a mix of vapor and liquid follow the contour map on the right side of each panel. The ``filled'' 2 nm models exhibit a thick liquid connection suspended between the upper and lower surfaces of the heat pipe. Additionally, a thin layer of liquid water can be observed closer to the inner surface of the heat pipe in both cases, which is unsurprising due to the strong copper-to-water attraction.

The ``medfill'' cases of the 2 nm model at both 350 K and 400 K do not display any unexpected features. However, for the ``medfill'' cases of the 3 nm model at both 400 K and 350 K, there are less dense regions at either end of the heat pipe. This occurs because the water molecules did not completely fill these regions during the initial equilibration stage, rather than due to evaporation-induced voids. Interestingly, the ``filled'' 3 nm model cases show complete wetting within the heat pipe and a relatively thicker layer near the inner surface. This layer contributes to the movement of water from the hot region to the cold region and is the primary contributor to heat transfer. This analysis will be further discussed later in this section, along with the mass flow rate and heat transfer rate estimation.

\subsection{Vorticity}

The vorticity $\omega$ in a 2D flow is defined as the curl of the velocity field. For a velocity field with components $v_x$ and $v_z$, the vorticity $\omega_y$ (since the system is periodic in y-axis, it's a 2D flow, and the vorticity will be perpendicular to the $x$-$z$ plane) is given by \cite{batchelor1988introduction, ferziger2002computational, hirsch1988numerical, springer2022fluid, physics2022vorticity}:

\begin{equation}
\omega_y = \frac{\partial v_z}{\partial x_g} - \frac{\partial v_x}{\partial z_g}
\end{equation}

In a discrete grid like the one we described earlier, the partial derivatives can be approximated using finite difference methods. For example, using central differences, the derivatives can be approximated as:

\begin{equation}
\frac{\partial v_z}{\partial x_g} \approx \frac{v_z(x_g + \Delta x_g, z_g) - v_z(x_g - \Delta x_g, z_g)}{2\Delta x_g}
\end{equation}

\begin{equation}
\frac{\partial v_x}{\partial z_g} \approx \frac{v_x(x_g, z_g + \Delta z_g) - v_x(x_g, z_g - \Delta z_g)}{2\Delta z_g}
\end{equation}

Thus, the discrete form of the vorticity can be written as:

\begin{equation}
\begin{aligned}
\omega_y(i, j) \approx \frac{v_z(i+1, j) - v_z(i-1, j)}{2\Delta x_g}  \\
- \frac{v_x(i, j+1) - v_x(i, j-1)}{2\Delta z_g}
\end{aligned}    
\end{equation}

\begin{figure*}[!h]
    \centering
    \includegraphics[width=\textwidth]{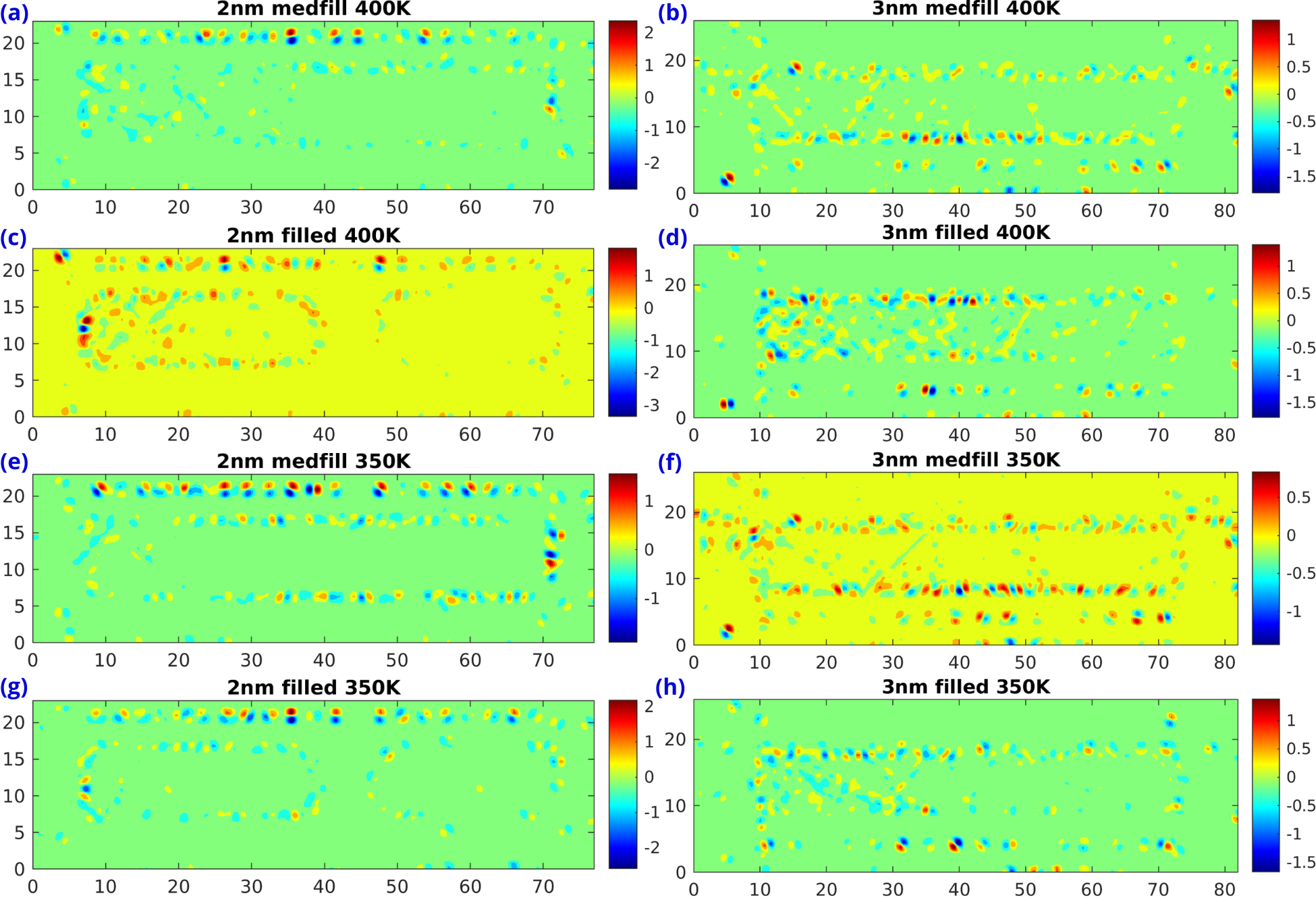}
    \caption{Vorticity plots for 2 nm model (1st column panels) and 3 nm model (2nd column panels) are given for high temperature (400 K) case in top four panels and low temperature (350 K) case in bottom four panels. Medium filled water cases are given in panels a, b, e, and f and filled water cases are given in c, d, g and h respectively.}
    \label{fig_vorticity}
\end{figure*}

Where $i$ and $j$ are the indices of the grid points in the $x_g$ and $z_g$ directions, respectively. The vorticity plots are shown in Fig. \ref{fig_vorticity}. The regions in red indicate areas of positive vorticity, which signifies that the fluid in these regions is rotating counterclockwise. Conversely, the regions in blue indicate areas of negative vorticity, where the fluid is rotating clockwise. For the 2 nm ''medfill'' cases (Fig. \ref{fig_vorticity}. a, e, and the high temperature filled case Fig. \ref{fig_vorticity}. g), the rotations are more localized inside the nanopores, particularly in the upper region. In the 2 nm filled high temperature case (Fig. \ref{fig_vorticity}. c), rotational motion is observed not only within the nanopore cavities but also along the interior surface of the heat pipe.

For all cases of the 3 nm model, rotation is predominantly seen along the inner surface of the heat pipe. This observation suggests that the flow of water occurs primarily at the surface of the heat pipe rather than through the nanopore wicks. The vorticity results for the 2 nm model do not indicate significant flow within the pores. Instead, these results imply that fluid energy is absorbed and transferred, facilitating local rotation. This claim is further corroborated by the velocity profile findings discussed next.

In summary, the vorticity distribution reveals distinct rotational behavior dependent on nanopore size and temperature conditions. The 2 nm model exhibits localized vorticity within the nanopores, particularly under higher temperature conditions, while the 3 nm model shows a surface-dominated rotational flow.

\begin{figure*}[!h]
    \centering
    \includegraphics[width=\textwidth]{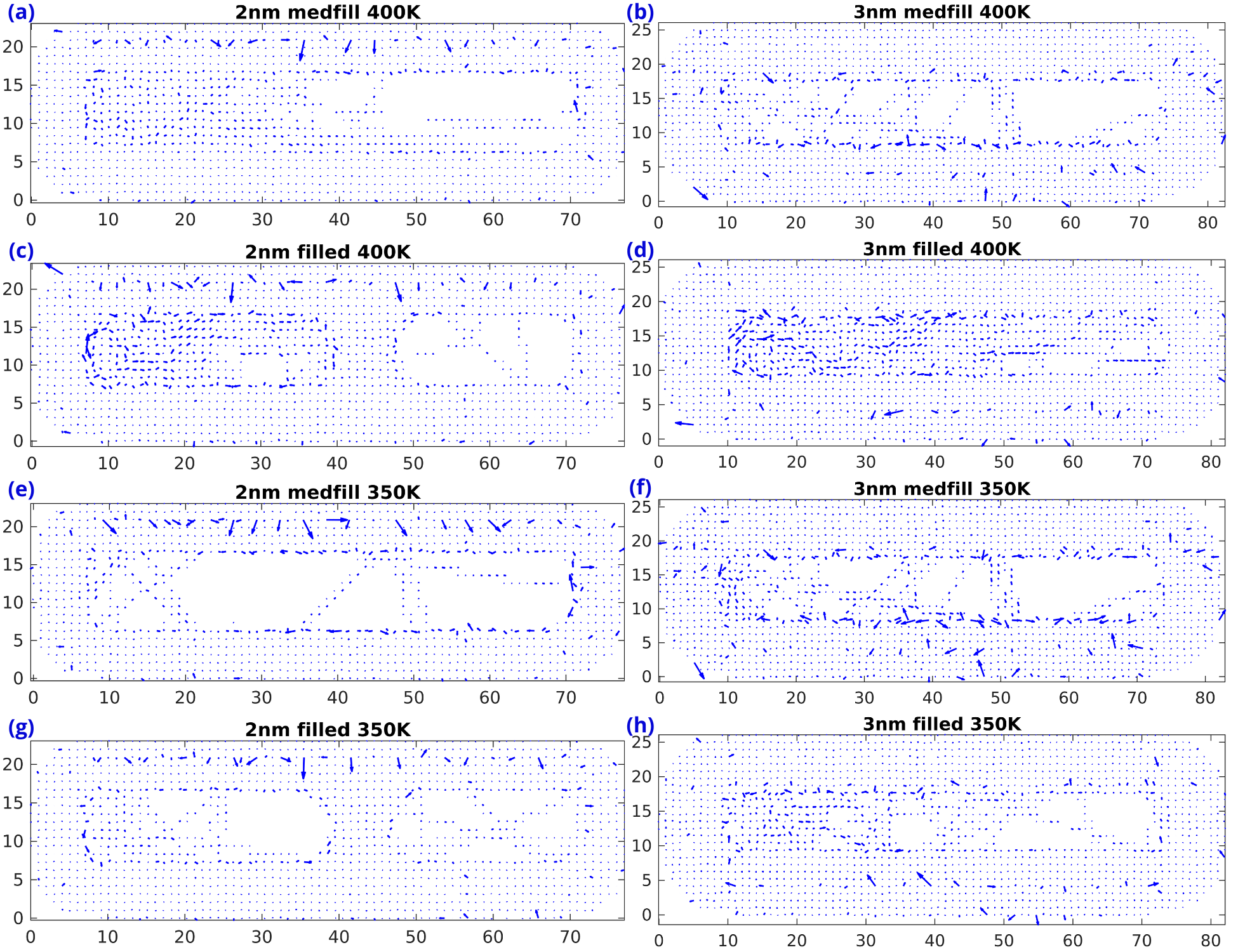}
    \caption{Velocity plots for 2 nm model (1st column panels) and 3 nm model (2nd column panels) are given for high temperature (400 K) case in top four panels and low temperature (350 K) case in bottom four panels. Medium filled water cases are given in panels a, b, e, and f and filled water cases are given in c, d, g and h respectively.}
    \label{fig_velocity}
\end{figure*}

The velocity plots for all the cases studied in this project are shown in Fig. \ref{fig_velocity}. The 2 nm medfill cases (Fig. \ref{fig_velocity} a and e) exhibit a continuous flow along the inner surface of the heat pipe, with stronger flows directed from the upper voids towards the lower regions. However, there is a lack of evidence suggesting a consistent pattern of heat transfer in these two cases.

For the 2 nm filled cases, the density plot indicates the formation of a thick water bridge between the upper and lower surfaces of the heat pipe. This bridge is sustained throughout the simulation and can be observed in the accompanying movie included in the supporting information. Figure \ref{fig_velocity} c) shows a strong recirculation of water along the inner surface in a clockwise direction. This recirculation is more pronounced at higher temperatures, particularly in the 400 K case, and weaker in the 350 K case (Fig. \ref{fig_velocity} g).

The 3 nm cases reveal that no significant lateral flows occur within the nanopores. Instead, the flow is predominantly along the inner surface of the heat pipe. Notably, the left half of the lower surface experiences a leftward flow, while the lower right half experiences a rightward flow. The high-temperature case of the filled 3 nm model suggests increased evaporation at the hot section of the heat pipe. A similar effect is observed in the 350 K case for the filled model, indicating enhanced evaporative cooling under elevated temperatures.

These observations provide insights into the fluid dynamics and heat transfer mechanisms within the nanoporous structures of the heat pipe, highlighting the influence of pore size and temperature on the overall performance. The sustained water bridge in the 2 nm cases suggests a stable liquid-vapor interface, which may contribute to localized heat transfer, while the 3 nm cases demonstrate surface-dominated flow patterns, emphasizing the importance of the inner surface in heat transport processes.

The velocity plot illustrates the arrow lengths corresponding to the magnitude, given by $|v|=\sqrt{v_x^2+v_z^2}$. This often poses challenges in visualizing the flow pattern within a region, particularly in areas with significantly lower velocities, such as the interior of the heat pipe wall compared to its surface. This visualization issue is evident in the plots shown in Fig. \ref{fig_velocity}, where the lower velocities inside the heat pipe wall make it difficult to discern the flow patterns.

\begin{figure*}[h]
    \centering
    \includegraphics[width=\textwidth]{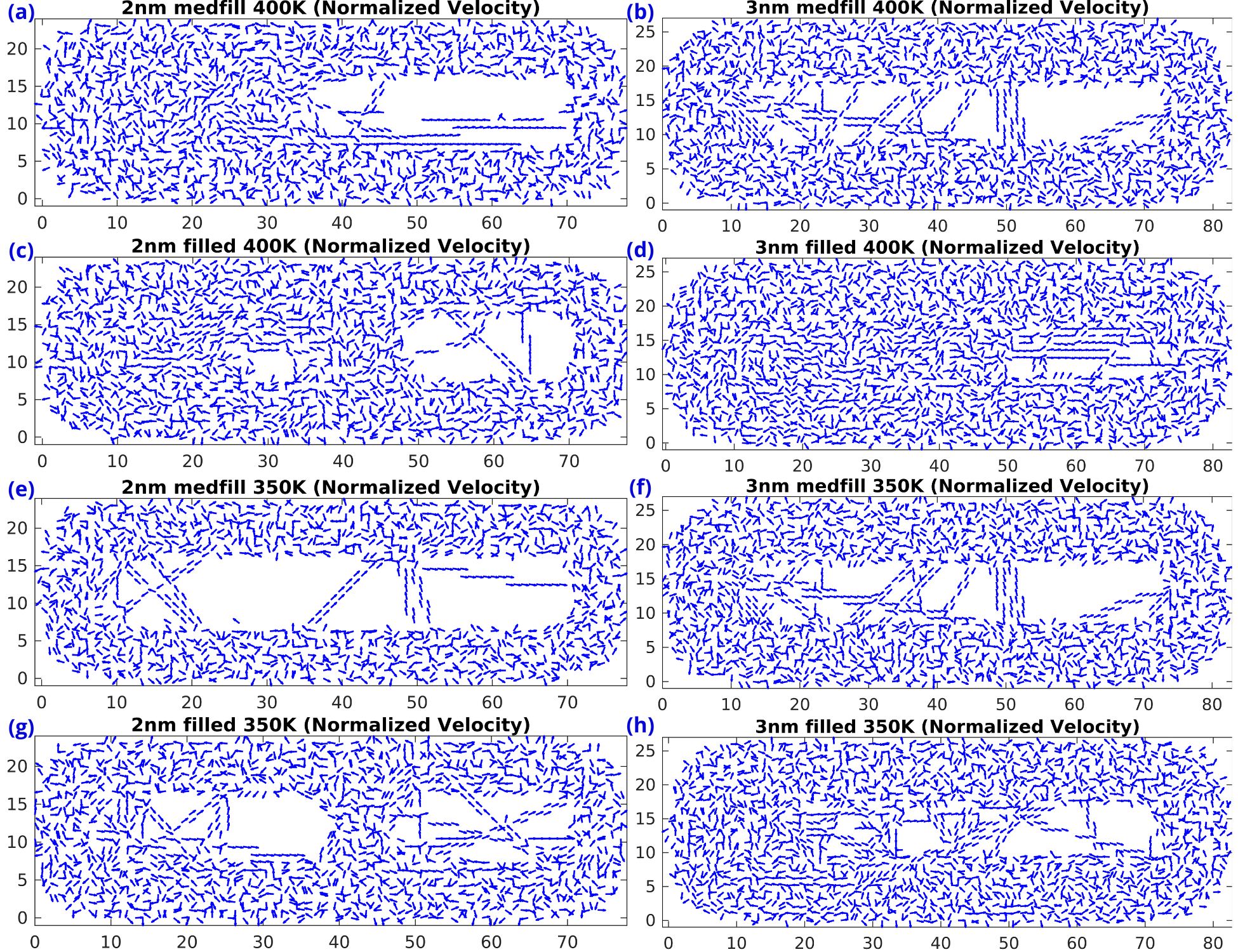}
    \caption{Normalized velocity plots for 2 nm model (1st column panels) and 3 nm model (2nd column panels) are given for high temperature (400 K) case in top four panels and low temperature (350 K) case in bottom four panels. Medium filled water cases are given in panels a, b, e, and f and filled water cases are given in c, d, g and h respectively.}
    \label{fig_normvelocity}
\end{figure*}

To address this issue, we can use normalized velocity vectors, where a constant length is applied to all velocity values. This normalization helps in better visualizing the flow patterns by ensuring that even regions with low velocities are adequately represented. The normalized velocity plot, which employs a constant length for all velocity vectors, is shown in Fig. \ref{fig_normvelocity}.

This approach allows for a clearer representation of the flow dynamics throughout the heat pipe, providing a better understanding of the fluid behavior across different regions. The normalized plots help in identifying flow patterns that might be overlooked in the standard velocity plots, thus offering a more accurate depiction of the fluid motion within the heat pipe.

\subsection{Mass Flow Rate}

The mass flow rate $\dot{m}$ $(kg/s)$ of the heat pipe is estimated by averaging the quantities over the entire 2D grid region, both temporally and spatially. This comprehensive approach ensures that the calculated mass flow rate accurately reflects the overall behavior of the heat pipe system under various conditions. The heat transfer rate $q$ $(W)$ can be estimated using the equation below:

\begin{equation}
q = \dot{m} c_p \Delta T
\end{equation}

Where $c_p$ is the specific heat of water $(4187 \ J/kg \cdot K)$ and $\Delta T$ is the temperature difference between the left and right ends of the heat pipe in Kelvin. This relationship underscores the direct dependence of the heat transfer rate on the mass flow rate, the specific heat capacity of the working fluid, and the temperature gradient across the heat pipe.

\begin{figure}[!h]
    \centering
    \includegraphics[width=0.7\textwidth]{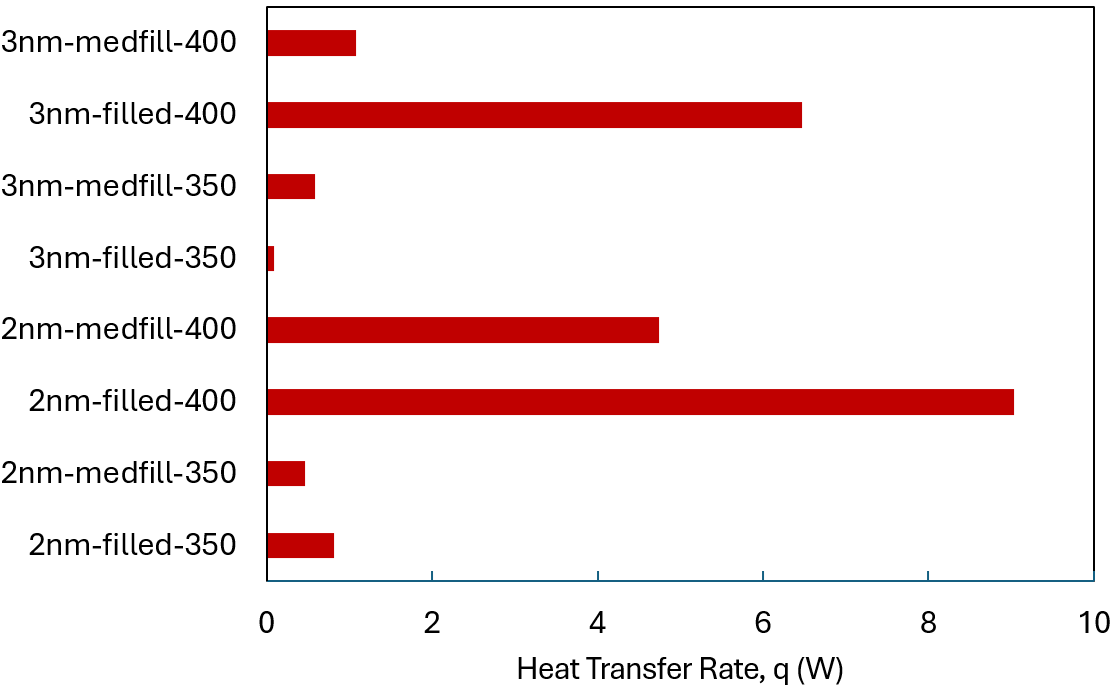}
    \caption{The heat transfer rate of the heat pipes under various conditions are shown in here. The x-axis represent heat transfer rate ($q$) and the y-axis represents various cases.}
    \label{fig_heatrate}
\end{figure}

In our analysis, the averaged quantities of mass flow rate and temperature difference are plotted in Fig. \ref{fig_heatrate}. The results indicate a clear trend where the cases at 400 K exhibit a superior heat transfer performance compared to those at 350 K. Specifically, the higher temperature gradient in the 400 K cases enhances the driving force for heat transfer, thereby increasing the overall efficiency of the heat pipe.

Additionally, the data reveal that the "filled" cases generally outperform the medium-filled cases in terms of heat transfer rate. This observation suggests that a higher fill ratio of the working fluid within the heat pipe improves the thermal performance, likely due to more effective phase change and fluid movement mechanisms.

Interestingly, the case with a 2 nm fill ratio surpassed the 3 nm model in terms of heat transfer rate performance. This seemingly counter intuitive result may be explained by an optimal balance between capillary action and fluid flow resistance at the 2 nm fill ratio, which maximizes the heat transfer efficiency.

In the case of the 2nm pores, capillary action does not significantly facilitate fluid flow within the pores themselves due to the small pore size. Instead, the primary mechanism responsible for the increased heat transfer rate is the surface-driven flow along the interior surfaces of the heat pipe. The strong temperature gradient between the hot and cold ends enhances this surface flow, as fluid movement is concentrated along the walls of the heat pipe rather than through the pores. 

It is important to note that the increased heat transfer rate observed in the 2nm model is not due to an inherent enhancement from the smaller pore size. Rather, the surface interactions, driven by the temperature gradient, dominate the fluid behavior and result in improved heat transfer. While capillary action contributes minimally in the 2nm pores, surface-driven flow plays the critical role, as the high surface-to-volume ratio amplifies these effects.

This finding underscores the importance of surface interactions in nanoscale heat pipes, where the flow dynamics shift away from traditional capillary-driven mechanisms and are instead governed by temperature-induced surface flows.

For filled high-temperature cases, we observed heat transfer rates ranging from $6.5 W$ to $9 W$, which are comparable to the performance of heat pipes used in laptops and computers. Typically, laptop heat pipes can handle power levels around $25 W$ to $52 W$ depending on their design and application parameters \cite{celsia2022heat, electronicscooling2022design, intech2022cooling, act2022heat}. Our findings underscore the importance of optimizing both the operating temperature and fill ratio in the design and application of heat pipes for efficient thermal management. These analysis and comparisons of these variables provide  insights into the thermal dynamics of heat pipes, guiding future improvements in their performance and application.

In calculating the heat transfer rate, we focused on the advective component using the equation
\[
q_{\text{adv}} = \dot{m} \, c_p \, \Delta T,
\]
where $\dot{m}$ is the mass flow rate, $c_p$ is the specific heat capacity of water, and $\Delta T$ is the temperature difference between the hot and cold ends. While this captures the heat carried by the moving fluid due to mass transport, we acknowledge that conductive heat transfer through the copper walls and the water also contributes to the overall heat transfer. To assess the impact of this omission, we estimated the conductive heat transfer using Fourier's Law:
\[
q_{\text{cond}} = -k_{\text{copper}} \, A \, \frac{\Delta T}{L},
\]
where $k_{\text{copper}}$ is the thermal conductivity of copper, $A$ is the cross-sectional area, and $L$ is the length over which conduction occurs. Our calculations showed that $q_{\text{cond}} \approx 92.5\, \mu\text{W}$, which is negligible compared to the advective heat transfer rates of $6.5$ to $9\,\text{W}$ observed in our simulations. This supports our focus on the advective term as the dominant heat transfer mechanism in this nanoscale system.


\subsection{Limitations and Scope for Improvements}

\subsubsection{Selection of Water Molecules}

Despite performing a comprehensive study on the system, we believe there are several factors that can be improved. The current study selected two levels of water with "filled" indicating water content visibly excess inside the heat pipe and "medfill" case with the bare minimum water for filling the nanopores. This selection is done through visual observation and can be improved systematically. The effect of water level on the heat transfer rate can be studied through a series of simulations of systems with varying levels of water, which will be computationally exhaustive. In fact, in the literature, there are no articles or technical documents explaining the accurate amount of water required based on a theoretical basis, which makes this problem a challenging one.

To address this challenge, future studies could adopt a systematic approach by incrementally varying the amount of water and performing detailed simulations for each level. This would provide a more accurate understanding of how the water content influences the heat transfer dynamics. Although this approach would require significant computational resources, it would offer valuable insights into optimizing the water level for enhanced thermal performance.

\subsubsection{Thermostatting Challenges and Alternatives}

The proper way of performing a heat pipe simulation using molecular dynamics is to thermostat the copper atoms and then leave the water molecules for NVE (constant number of particles, volume, and energy) ensemble integration. Though in theory, this approach looks accurate, practically it creates an unstable system even within the lower ends of acceptable time steps of integration for coarse-grained molecular dynamics. An improvement in this context could be dividing the copper into tiny sections (probably 100+) and then thermostating the water within each section. This would require the remaining water within the inside of the heat pipe to be NVE integrated.

Although this approach is feasible, it can significantly slow down the computational process due to the need for frequent region updates. This is an area that could be explored further to develop more stable and efficient simulation techniques. Advanced thermostatting methods or hybrid approaches combining different ensembles could potentially mitigate these stability issues while maintaining computational efficiency.

\subsubsection{Ideal Length of Heat Pipe}

There is no universally ideal length for a heat pipe; instead, it is often derived from the design and physical needs of the system under consideration. While the length can influence the mass flow rate and heat transfer rate, it is less likely to change the mode of heat transfer. Most dynamics settle down within the middle section of the heat pipe, suggesting that our model length is sufficient for this study.

However, future sensitivity studies could be performed to understand the effect of heat pipe length on various thermodynamic parameters. By varying the length and observing the resulting changes in performance, researchers can derive more precise guidelines for optimizing heat pipe dimensions in practical applications.

\subsubsection{Separation of Velocity Effects in Temperature and Pressure Estimation}

One major observation while simulating the system with sub-critical nanopores is the bulk movement of water within the heat pipe. This bulk velocity can affect the calculation of temperature and pressure, as shown in Equations \ref{eq_temp} and \ref{eq_press}.

\begin{equation}
T = \frac{1}{3 N k_B} \sum_{i=1}^{N} m_i v_i^2
\label{eq_temp}
\end{equation}

\begin{equation}
P = \frac{1}{V} \left( N k_B T + \frac{1}{3} \sum_{i=1}^{N} \sum_{j>i}^{N} \mathbf{r}{ij} \cdot \mathbf{f}{ij} \right)
\label{eq_press}
\end{equation}

We attempted to correct for this by removing the center of mass velocity from the system and also by removing the bulk velocity for each grid point. However, this strategy did not significantly improve the results and its limitations can be observed in the temperature plots provided in the supplementary document.

Future work could explore more sophisticated techniques for separating the bulk velocity effects from the intrinsic thermal motions. For instance, advanced filtering methods or improved algorithms for velocity decomposition might provide better accuracy in estimating the true thermodynamic properties of the system.

One limitation of our current model is the exclusion of explicit heat transfer between the copper walls and the working fluid. In conventional capillary heat pipes (CHPs), thermal interactions between the wall, wick, and working fluid are significant and can influence overall performance. By not directly thermostating the copper atoms, our model simplifies the system to focus on the fluid behavior within the nanopores. Future work could involve coupling the temperature of the copper walls with the fluid dynamics to provide a more comprehensive simulation that captures the heat transfer between all components of the heat pipe, thereby offering insights that are more directly comparable to larger-scale CHPs.

\section{Conclusion}

This study offers a detailed investigation into the thermal dynamics of heat pipes with sub-critical nanopores using coarse-grained molecular dynamics (CGMD) simulations. With the growing demand for efficient thermal management in microelectronics and high-performance computing, heat pipes with nanoporous structures have emerged as promising candidates due to their high thermal conductivity and passive operation. Traditional heat pipes rely on capillary action within wicking structures, but recent advancements in nanotechnology have enabled the development of nanoporous heat pipes that enhance heat transfer capabilities at micro and nanoscale levels. Our research focuses on modeling water molecules and copper structures at the nanoscale to elucidate key factors affecting heat transfer efficiency in these systems.

The results reveal that the heat transfer rate is significantly higher at larger temperature differences, such as 400 K compared to 350 K, due to the increased driving force for thermal energy transport. Additionally, cases with higher water fill ratios demonstrate superior thermal performance, attributed to more effective phase change processes and fluid dynamics within the heat pipe. Interestingly, the 2 nm filled models show better heat transfer performance than the 3 nm filled models, suggesting an optimal balance between capillary action and fluid flow resistance at the 2 nm fill ratio. Vorticity and velocity analyses highlight distinct flow dynamics; the 2 nm models exhibit localized vorticity within the nanopores, especially at higher temperatures, while the 3 nm models demonstrate surface-dominated rotational flow, indicating that water flow primarily occurs along the inner surfaces of the heat pipe. These findings suggest that in heat pipes with sub-critical nanopores, surface-driven flows are more significant than traditional wicking action, providing crucial insights for optimizing nanopore size and fill ratios to enhance heat transfer efficiency. This study advances our understanding of the thermal behavior of nanoporous heat pipes and  contributing to the development of more efficient thermal management solutions in various applications, particularly in electronics cooling and energy systems.

\section*{Acknowledgment}

The author gratefully acknowledges the University of New Haven for providing startup funding that facilitated the acquisition of workstations essential for this research.

\bibliographystyle{elsarticle-harv}   
\bibliography{REFERENCES}            

\end{document}